\documentclass{aa} 

\usepackage{graphicx}
\usepackage{txfonts}
\usepackage{xcolor}

\begin{document}

   \title{A high-resolution radio study of the L1551 IRS 5 and L1551 NE jets}

   \author{A. Feeney-Johansson
          \inst{1,2}
          \and
          S. J. D. Purser
          \inst{3}
          \and
          T. P. Ray
          \inst{1}
          \and
          C. Carrasco-González
          \inst{4}
          \and
          A. Rodríguez-Kamenetzky
          \inst{5,6}
          \and
          J. Eisl\"offel
          \inst{7}
          \and
          J. Lim
          \inst{8}
          \and
          R. Galván-Madrid
          \inst{4}
          \and
          S. Lizano
          \inst{4}
          \and
          L. F. Rodríguez
          \inst{4}
          \and
          H. Shang
          \inst{9}
          \and
          P. Ho
          \inst{9,10}
          \and
          M. Hoare
          \inst{11}
          }

   \institute{Dublin Institute for Advanced Studies, Astronomy \& Astrophysics Section, 31 Fitzwilliam Place, Dublin, D02 XF86, Ireland\\
              \email{antonfj@astron.s.u-tokyo.ac.jp}
        \and
            National Astronomical Observatory of Japan, Osawa 2-21-1, Mitaka, Tokyo 181-8588, Japan
        \and
            SKA Observatory, Jodrell Bank, Lower Withington, Macclesfield, Cheshire, SK11 9FT, UK 
        \and
            Instituto de Radioastronomía y Astrofísica (IRyA-UNAM), Morelia, Mexico
        \and
             Instituto de Astronomía Teórica y Experimental (IATE, CONICET-UNC), Laprida 854, Córdoba, X5000BGR, Argentina
        \and
            Observatorio Astronómico de Córdoba, Universidad Nacional de Córdoba, Laprida 854, X5000BGR, Córdoba, Argentina.
        \and
            Thüringer Landessternwarte, Sternwarte 5, D-07778 Tautenburg, Germany
        \and
            Department of Physics, The University of Hong Kong, Pokfulam Road, Hong Kong
        \and
            Institute of Astronomy and Astrophysics, Academia Sinica, Taipei 10617, Taiwan
        \and
            East Asian Observatory, 660 N. A’ohōkū Place, Hilo, Hawaii, HI 96720, USA
        \and
            School of Physics and Astronomy, University of Leeds, Leeds, UK
        }

   \date{}

 
  \abstract{Using observations with e-MERLIN and the VLA, together with archival data from ALMA, we obtain high-resolution radio images of two binary YSOs: L1551 IRS 5 and L1551 NE, covering a wide range of frequencies from 5 - 336 GHz, and resolving emission from the radio jet on scales of only $\sim 15\ \mathrm{au}$. By comparing these observations to those from a previous epoch, it is shown that there is a high degree of variability in the free-free emission from the jets of these sources. In particular, the northern component of L1551 IRS 5 shows a remarkable decline in flux density of a factor of $\sim 5$, suggesting that the free-free emission of this source has almost disappeared. By fitting the spectra of the sources, the ionised mass-loss rates of the jets are derived and it is shown that there is significant variability of up to a factor of $\sim 6$ on timescales of $\sim 20$ years. Using radiative transfer modelling, we also obtained a model image for the jet of the southern component of L1551 IRS 5 to help study the inner region of the ionised high-density jet. The findings favour the X-wind model launched from a very small innermost region.}


   \maketitle
%

\section{Introduction}
Jets and outflows are observed towards young stellar objects (YSOs) ranging in mass from proto-brown dwarfs to O-type protostars, and play an important role in star formation \citep{Frank2014}. They are closely linked to accretion and help extract angular momentum from infalling matter thus permitting accretion onto the protostar \citep{Bacciotti2002}.

Observations at radio wavelengths permit studies of protostellar outflows close to their driving sources, where observations at optical and near-infrared wavelengths are difficult due obscuration by dust. Free-free (bremsstrahlung) emission at cm-wavelengths is commonly observed towards the base of the jet and is thought to trace material which has been ionised by internal shocks in the jet \citep{Curiel1987}. Radio observations at cm-wavelengths can therefore help to trace material which has recently been ejected by the YSO and calculate properties such as the direction, collimation, or mass-loss rate of this recently ejected material. This can be compared with observations which track the larger-scale outflow, such as molecular outflows or optical/IR jets.

In spite of many observational studies of protostellar outflows, the mechanism by which they are launched and collimated is still broadly debated \citep{Ray2021}. It is thought that material is accelerated by magnetocentrifugal forces along the magnetic field lines of either the star or disk. The two main theories are known as the X-wind model \citep{Shu1994} and the disk wind model \citep{Blandford1982,Pudritz1983}. In the X-wind model, material is thought to be launched from the innermost region of the disk
along the pinched stellar magnetic field lines. In the disk wind model, material is launched from the surface of the disk at a wide range of distances from the star up to tens of au along open magnetic field lines threading the disk. In both models, the material is then ``self-collimated'' far beyond the Alfv\'en radius through magnetic hoop stresses at large scale.

In order to learn more about the launching and collimation of jets, high resolution observations near the base of the jet are necessary. Optical or IR emission from this region is often obscured by high levels of extinction, particularly in more embedded sources. Therefore, radio observations of the free-free emission are ideally suited for this purpose. Previous observations at cm-wavelengths indicate that jets are collimated on scales of only a few tens of au or less \citep{Anglada2018}. The highest resolution observations of a low-mass protostar to date \citep{Carrasco-Gonzalez2019}, suggest that in HL Tau the jet is already collimated at only $\sim 1.5\ \mathrm{au}$ from the star. On the other hand, in the high-mass YSO Cep A HW2, \citet{Carrasco-Gonzalez2021} found that the jet was collimated further away from the protostar than in the case of low-mass YSOs, at $\sim 20 - 30\ \mathrm{au}$. They propose two possible launching mechanisms to explain this, either a scaled up version of the disk wind model, where the collimation distance scales with the mass of the star, or a wide-angle wind launched from the disk which is then externally collimated at large distances from the star by a large-scale magnetic field or dense ambient medium. In the intermediate-mass YSO Serpens SMM1, \citet{Rodriguez-Kamenetzky2022} suggest that its outflow consists of two components, a highly collimated narrow jet launched by the X-wind mechanism and a wide-angle wind which is launched by either the X-wind or disk wind mechanism.

In this work, we report on new high-resolution observations of two YSO systems in the L1551 star forming region: L1551 IRS 5 and L1551 NE. L1551 is a low-mass star forming region located in the south of the Taurus Molecular Cloud (TMC) at a distance of $147\pm5\ \mathrm{pc}$ \citep{Connelley2018}. The nearby location of this region makes it an ideal target for high-resolution observations as it allows us to resolve the regions close to the star. These observations were carried out using e-MERLIN and the Very Large Array (VLA) at a range of frequencies from 5 - 26 GHz. The resolution of these observations allows us to discern emission on scales of only $\sim 15\ \mathrm{au}$.

\begin{table*}
\centering
\caption{Summary of Observations of L1551 IRS 5}
\begin{tabular}{c c c c c c c}
	\hline
	Instrument  & Date          & Time      & $\lambda $    & $\nu$ & $\Delta \nu $ & TOS \\
	            &               &           & (mm)          & (GHz) & (GHz)         & (min) \\
	\hline
	e-MERLIN    & 2020-Jan-25   & 15:32:05  & 60            & 5     & 0.5           & 990 \\
	VLA         & 2021-Jan-18   & 06:34:54  & 50            & 6     & 4             & 6     \\
	VLA         & 2021-Jan-15   & 06:07:40  & 30            & 10    & 4             & 6     \\
	VLA         & 2021-Jan-09   & 00:38:00  & 20            & 15    & 6             & 80    \\
	VLA         & 2020-Dec-29   & 02:41:00  & 14            & 22    & 8             & 80  \\
	ALMA        & 2017-Nov-20   & 04:05:30  & 3.2           & 93    & 8             & 4      \\
	ALMA        & 2017-Nov-04   & 07:34:18  & 2.0           & 153   & 8             & 6      \\
	ALMA        & 2017-Jul-24   & 12:23:52  & 1.3           & 225   & 4             & 11 \\
	ALMA        & 2017-Jul-27   & 11:32:46  & 0.9           & 336   & 6             & 36 \\
	\hline
\end{tabular}
\label{tab:observations_L1551_IRS_5}
\end{table*}

\begin{table*}
\centering
\caption{Summary of Observations of L1551 NE}
\begin{tabular}{c c c c c c c}
	\hline
	Instrument  & Date          & Time      & $\lambda$ & $\nu$ & $\Delta \nu$  & TOS   \\
	            &               &           & (mm)      & (GHz) & (GHz)         & (min) \\
	\hline
	VLA         & 2021-Jan-18   & 06:34:54  & 50        & 6     & 4             & 6     \\
	VLA         & 2021-Jan-15   & 06:07:40  & 30        & 10    & 4             & 6     \\
	VLA         & 2021-Jan-09   & 00:38:00  & 20        & 15    & 6             & 80    \\
	VLA         & 2020-Dec-29   & 02:41:00  & 14        & 22    & 8             & 80    \\
	ALMA        & 2019-Oct-07   & 06:47:31  & 0.9       & 336   & 6             & 18 \\
	\hline
\end{tabular}
\label{tab:observations_L1551_NE}
\end{table*}

L1551 IRS 5 is a well-known Class I source which has been classified as a FUor object based on spectral analysis \citep{Connelley2018}. These are objects which show strong outbursts, in which their brightness increases by several orders of magnitude, and are thought to be caused by rapid increases in the accretion rate of the star. Such outbursts can then last for decades \citep{Audard2014}.  L1551 IRS 5 is a binary with a projected separation of $\sim0.36\arcsec$ or $\sim 50$ au between the northern (N) and southern (S) components. The total mass of the system is estimated to be $\sim 0.5 - 1.0\ \mathrm{M_{\odot}}$ with a mass ratio between the two components of $q \sim 1$ \citep{Rodriguez2003b,Lim2006,Chou2014}. Each has a circumstellar disk as well as a circumbinary disk surrounding the whole system \citep{Cruz-SaenzDeMiera2019,Takakuwa2020}. Both components have high mass accretion rates of $\dot{M}_{\mathrm{acc}} = 6 \times 10^{-6}\ \mathrm{M_{\sun}\ yr^{-1}}$ and $\dot{M}_{\mathrm{acc}} = 2 \times 10^{-6}\ \mathrm{M_{\sun}\ yr^{-1}}$ for the N and S components respectively \citep{Liseau2005}. Both of the components are known to drive jets which are seen in the form of thermal ionised jet emission in the radio \citep{Rodriguez2003} as well as in the optical \citep{Mundt1983}, and are collimated within $\leq 3\ \mathrm{au}$ from the central star \citep{Lim2006}. The N jet has a position angle of $67\pm3 \degr$ while the S jet has a position angle of $55\pm1 \degr$ with the red-shifted portions of both jets pointing to the NE and the blue-shifted portions to the SW\citep{Rodriguez2003}. L1551 IRS 5 was also the first source towards which a molecular outflow was detected \citep{Snell1980}. The projection angle for both circumstellar disks was estimated by \citet{Lim2016a} to be $\sim 45 \degr$ based on VLA mm observations of the dust emission. From this, we can assume the inclination angle for both jets to the line of sight is also approximately $i \approx 45 \degr$. Based on [\ion{O}{I}]$_{63}$ observations of the atomic jet and by comparison with other atomic and molecular emission lines, \citet{Sperling2021} determine that the atomic jet is the dominant component of the outflow and estimate the mass-loss rate of the outflow to be $\dot{M} = (4.9 - 5.4 \pm 0.8) \times 10^{-7}\ \mathrm{M_{\sun}\ yr^{-1}}$.

L1551 NE is another Class I binary system located $\sim2.5\arcmin$, or $\sim 22000$ au, northeast of L1551 IRS 5. It is comprised of a southeastern source (known as component A) and a northwestern source (known as component B) with an estimated projected separation between the two components of $\sim0.5\arcsec$ or $\sim 70$ au. The deprojected separation between the two components is calculated to be $\sim 145$ au \citep{Takakuwa2014}. The total mass of the system is estimated to be $\sim 0.8\ \mathrm{M_{\odot}}$ with a mass ratio of $M_{\mathrm{B}} / M_{\mathrm{A}} \sim 0.19$, where $M_{\mathrm{A}}$ and $M_{\mathrm{B}}$ are the mass of component A and B respectively \citep{Takakuwa2012}. Similar to L1551 IRS 5, each component has a circumstellar disk as well as a circumbinary disk surrounding the whole system \citep{Takakuwa2014,Takakuwa2017}. Thermal ionised jet emission is also seen from both components with position angles for the jets of $48\pm10 \degr$ and $58\pm8 \degr$ for components A and B respectively with the red-shifted portions of both jets pointing to the NE and the blue-shifted portions to the SW \citep{Reipurth2002}. The outflow from  L1551 NE is thought to be responsible for the HH objects HH 28 and HH 29 and a series of HH knots known as HH 454 which surrounds L1551 NE. \citet{Lim2016b} calculated the projection angles of both disks, and hence the inclination angles of both jets to the line of sight, to be $\sim 58\degr$ based on VLA mm observations. From this, the inclination angle of the jet to the line of sight is estimated to also be $i \approx 58\degr$.

The structure of this paper is as follows. In \autoref{sec:observations}, we describe the radio observations carried out using e-MERLIN, the VLA, and ALMA and the data reduction process. In \autoref{sec:results}, we present the resulting images obtained and the flux densities measured at the different frequencies. In \autoref{sec:discussion}, we then model the spectra of the sources. This was then used to estimate the ionised mass-loss rate for each jet. The variability seen in the jets compared to previous observations is also discussed. Radiative transfer modelling is also performed for one of the jets in L1551 IRS 5. Finally, in \autoref{sec:conclusions}, we present our concluding remarks.

\section{Observations and Data Reduction}
\label{sec:observations}
L1551 IRS 5 was observed using e-MERLIN, including the Lovell Mk I 76 m telescope, on 2020 January 25 in C Band at 5 GHz/6 cm with a total bandwidth of 0.5 GHz (Project code: LE1007). The total time-on-source was $\approx$16 hrs. Using the VLA in the A configuration, the most extended configuration for the array, L1551 IRS 5 and L1551 NE were observed at C Band (6 GHz/5 cm), X Band (10 GHz/3 cm), Ku Band (15 GHz/2 cm), and K Band (22 GHz/1.4 cm) on 2021 January 18, 15, 19, and 2020 December 29, respectively (Project code: 20B-122). All of the observations were calibrated and imaged using the Common Astronomy Software Applications \citep[CASA;][]{CASA2022} software package (version 6.1.2). For the e-MERLIN observation, 3C286 was used as the flux calibrator, J1407+2827 was used as the bandpass calibrator, and J0428+1732 was used as the complex gain calibrator. For the VLA observations, 3C147 was used as the flux calibrator at all frequencies and was also used as the bandpass calibrator for the C Band and X Band observations. For the VLA Ku Band and K Band observations, 3C84 was used as the band pass calibrator.  The source J0431+2037 was used as the complex gain calibrator in all VLA observations for both L1551 IRS 5 and L1551 NE. Self-calibration was also successfully performed for the VLA Ku band and K band observations, although it was not possible for the e-MERLIN observation nor the VLA C band and X band observations due to insufficient signal-to-noise.

As well as these observations we also used archival data for our sources from the Atacama Large Millimeter/submillimeter Array (ALMA). L1551 IRS 5 was observed in Band 3 (93 GHz/3.2 mm) and Band 4 (153 GHz/2 mm) on 2017 November 20 and 2017 November 04 (Project code: 2017.1.00388.S; PI: H. I. Liu), in Band 6 (225 GHz/1 mm) on 2017 July 24 (Project code: 2016.1.00209.S; PI: M. Takami), and in Band 7 (336 GHz/0.9 mm) on 2017 July 27 (Project code: 2016.1.00138.S; PI: S. Takakuwa). L1551 NE was observed in Band 7 on 2019 October 07 (Project code: 2019.1.00847.S; PI: P. Sheehan). The Band 6 and Band 7 data for L1551 IRS 5 were previously published in \citet{Cruz-SaenzDeMiera2019} and \citet{Takakuwa2020} respectively. The archival ALMA images were inspected, along with the corresponding calibration inspections plots from the archive, to ensure that the data was of sufficiently high quality to use for our study. Details of all of the observations used in this study are given in \autoref{tab:observations_L1551_IRS_5} and \autoref{tab:observations_L1551_NE}.

\begin{table*}[t]
    \centering
    \caption{Flux density measurements in L1551 IRS 5 N component and S component}
    \begin{tabular}{ccccc}
        \hline 
        $\nu$ & L1551 IRS 5 N & L1551 IRS 5 S & $\sigma_{\mathrm{rms}}$ & Synthesised Beam \\
        (GHz) & (mJy) & (mJy) & $\mathrm{\mu Jy\ beam^{-1}}$ & \\
        \hline 
        5 & $0.11 \pm 0.02$ & $1.16 \pm 0.07$ & 11 & $0\farcs22 \times 0\farcs11$, $-44\degr$ \\ 
        10 & $0.47 \pm 0.03$ & $1.8 \pm 0.1$ & 10 & $0\farcs20 \times 0\farcs18$, $49\degr$  \\
        13.5 & $0.88 \pm 0.05$ & $2.0 \pm 0.1$ & 6 & $0\farcs17 \times 0\farcs13$, $-62\degr$ \\ 
        17.5 & $1.15 \pm 0.06$ & $2.1 \pm 0.1$ & 6 & $0\farcs14 \times 0\farcs11$, $-59\degr$ \\
        20 & $1.23 \pm 0.07$ & $2.1 \pm 0.1$ & 12 &  $0\farcs12 \times 0\farcs09$, $-18\degr$ \\
        24 & $1.60 \pm 0.09$ & $2.1 \pm 0.1$ & 12 & $0\farcs094 \times 0\farcs074$, $-29\degr$\\ 
        93 & $32 \pm 2$ & $18 \pm 1$ & 50 & $0\farcs124 \times 0\farcs089$, $19\degr$ \\
        153 & $79 \pm 4$ & $43 \pm 2$ & 76 & $0\farcs058 \times 0\farcs041$, $-33\degr$ \\ 
        225 & $240 \pm 10$ & $125 \pm 7$ & 750 & $0\farcs15 \times 0\farcs13$, $-85\degr$ \\
        336 & $420 \pm 20$ & $210 \pm 10$ & 1100& $0\farcs12 \times 0\farcs10$, $-65\degr$ \\
        \hline 
    \end{tabular}

    \label{tab:L1551_IRS_5_fluxes}
\end{table*}

\begin{table*}[t]
    \centering
    \caption{Flux density measurements in L1551 NE component A and component B}
    \begin{tabular}{ccccc}
        \hline 
        $\nu$ & L1551 NE A & L1551 NE B & $\sigma_{\mathrm{rms}}$ & Synthesised Beam \\
        (GHz) & (mJy) & (mJy) & ($\mathrm{\mu Jy\ beam^{-1}}$) & \\
        \hline 
        6 & $0.18 \pm 0.03$ & $0.16 \pm 0.02$ & 16 & $0\farcs37 \times 0\farcs30$, $61\degr$ \\
        10 & $0.31 \pm 0.04$ & $0.15 \pm 0.03$ & 26 & $0\farcs20 \times 0\farcs18$, $49\degr$ \\
        13.5 & $0.58 \pm 0.04$ & $0.27 \pm 0.02$ & 6 & $0\farcs17 \times 0\farcs13$, $-61\degr$ \\
        17.5 & $0.81 \pm 0.05$ & $0.33 \pm 0.02$ & 6 & $0\farcs14 \times 0\farcs11$, $-59\degr$ \\
        20 & $1.11 \pm 0.07$ & $0.43 \pm 0.03$ & 11 &  $0\farcs10 \times 0\farcs09$, $-14\degr$ \\
        24 & $1.36 \pm 0.08$ & $0.51 \pm 0.04$ & 11 &  $0\farcs084 \times 0\farcs073$, $-27\degr$ \\
        336 & $330 \pm 20$ & $132 \pm 8$ & 1100 & $0\farcs30 \times 0\farcs25$, $-12\degr$ \\
        \hline 
    \end{tabular}

    \label{tab:L1551_NE_fluxes}
\end{table*}

In order to achieve good uv-coverage and create a high resolution image of L1551 IRS~5 at C Band, the visibility data from the e-MERLIN C band observation and the VLA C band observation were combined. This takes advantage of the high resolution of e-MERLIN and the sensitivity and short baseline coverage of the VLA. As the observations are taken approximately a year apart, the proper motion of L1551 IRS~5 must be accounted for, as well as any phase errors between the two observations. The two observations use different phase calibrators, and so there may be differences in the positions of sources as a result. The absolute proper motion of the system was calculated by \citet{Villa2017} based on VLA observations at 7 mm and found to be on the order of $\mu \sim 25\ \mathrm{mas\ yr^{-1}}$ ($\mu_{\alpha} = +17.4 \pm 4.0\ \mathrm{mas\ yr^{-1}}$, $\mu_{\delta} = -18.2 \pm 0.6\ \mathrm{mas\ yr^{-1}}$). To account for any differences in position due to phase errors, the position of the point-like field source XEST 22-054, located at the co-ordinates (J2000) RA: 04:31:43.00 and Dec: +18:10:34.5, was compared between the two images. The source was assumed to be an extragalactic source based on its steep negative spectral index ($\alpha = -2.02 \pm 0.06$) in the VLA C Band image and the lack of any IR detections at this position when checked using SIMBAD. Therefore it should have no significant proper motion. The difference in the RA and Dec was found to be $\Delta_{\alpha} = +20\ \mathrm{mas}$ and $\Delta_{\delta} = -70\ \mathrm{mas}$ respectively. When combining the proper motion and the difference due to phase errors, this gives a total difference in position between the two observations of $\Delta_{\alpha} = +37.1\ \mathrm{mas}$ and $\Delta_{\delta} = -87.8\ \mathrm{mas}$. Therefore, the e-MERLIN data was shifted so that the position of L1551 IRS 5 would match the position in the VLA data using the CASA task \emph{fixvis}. It should be noted that there would also be proper motion of the emission knots in the jet relative to the star and disk, which cannot be corrected for since the knots in the red-shifted and blue-shifted portions would be moving in opposite directions. There is also relative proper motion of the two components with respect to each other due to orbital motion which also could not be corrected for, with the change in RA and Dec of the S component with respect to the N component being $\Delta_{\alpha} = -3.2 \pm 0.9\ \mathrm{mas\ yr^{-1}}$ and $\Delta_{\delta} =  -2.6 \pm 0.7\ \mathrm{mas\ yr^{-1}}$ \citep{Lim2006,Lim2016a}.  Since the e-MERLIN and VLA observations have different bandwidths, only the part of the VLA bandwidth corresponding to the e-MERLIN bandwidth was used so that they would have the same central frequency and bandwidth when combined into one set of data.

\begin{figure*}[h!]
\centering
	\includegraphics{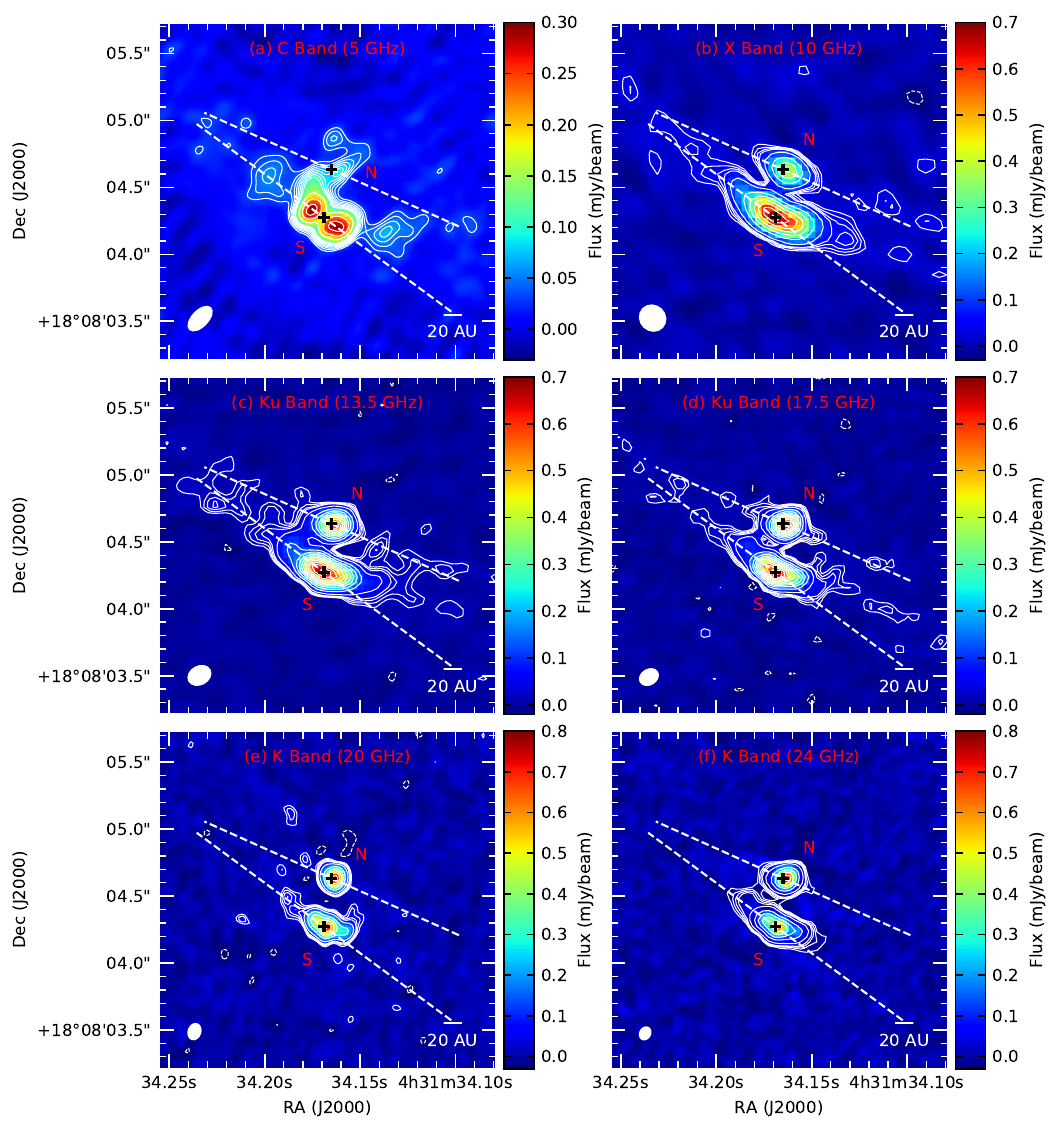}[p]
	\caption{Images of L1551 IRS 5 from the VLA and e-MERLIN observations (contours + colour scale) as follows: (a) the e-MERLIN + VLA 5 GHz image (b) the VLA 10 GHz image (c) the VLA 13.5 GHz image (d) the VLA 17.5 GHz image (e) the VLA 20 GHz image (f) the VLA 24 GHz image. The black crosses in each image indicate the peaks of the disk emission of the two sources in the ALMA 153 GHz image to show the approximate positions of the central stars. The jet axes of the two components from \protect{\citet{Rodriguez2003}} are shown by the dotted white lines. The contour levels in the 5 GHz image are -3, 3, 4, 5, 6, 7, 8, 9, 10, 15, 20, and 25 $\times\ \sigma_{rms}$, where $\sigma_{rms}$ is the root-mean-square noise of the image. For the 10 GHz, 20 GHz, and 24 GHz images, the contour levels are -3, 3, 4, 5, 6, 10, 20, 30, 40, 50, and 60 $\times\ \sigma_{rms}$. For the 13.5 GHz and 17.5 GHz images, the contour levels are -3, 3, 4, 5, 6, 10, 20, 30, 40, 50, 60, 70, 80, 90, 100, and 110 $\times\ \sigma_{rms}$. The synthesised beam for each image is shown in the bottom left corner. The values for the synthesised beam and $\sigma_{rms}$ of each image are given in \protect{\autoref{tab:L1551_IRS_5_fluxes}}.}
	\label{fig:L1551_IRS_5_C_to_K_Band}
\end{figure*}

\begin{figure*}
    \centering
    \includegraphics{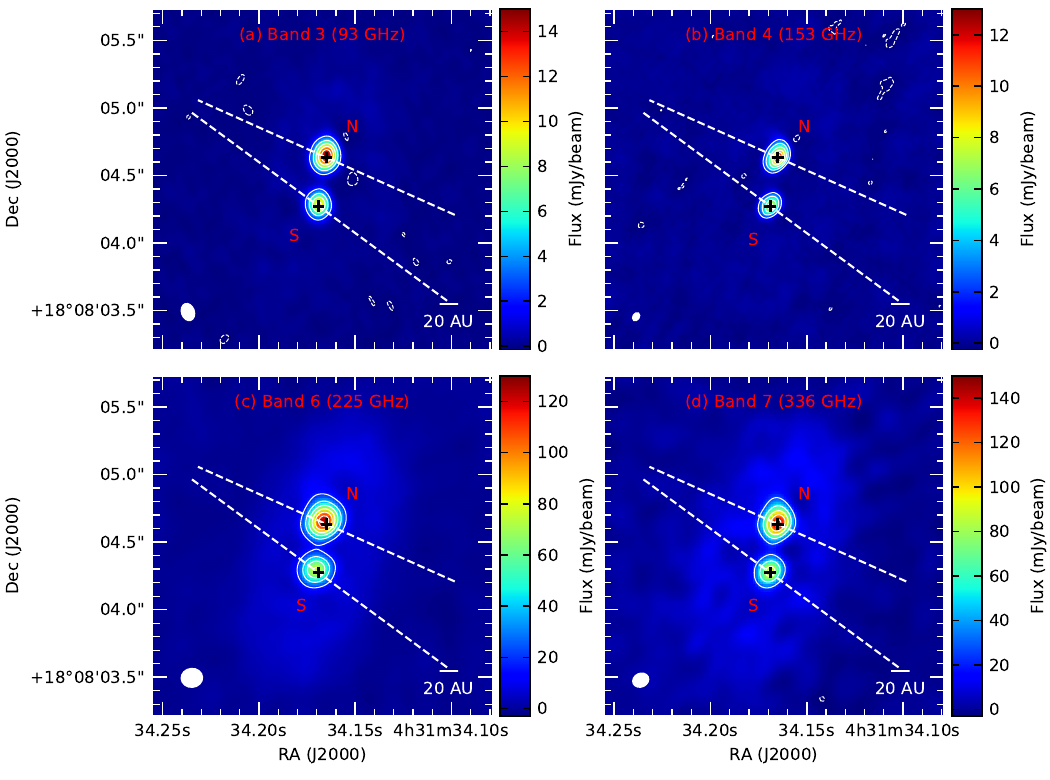}
    \caption{Images of L1551 IRS 5 from the ALMA observations (contours + colour scale) as follows: (a) the ALMA 93 GHz image (b) the ALMA 153 GHz image (c) the ALMA 225 GHz image (d) the ALMA 336 GHz image. The black crosses in each image indicate the peaks of the disk emission of the two sources in the ALMA 153 GHz image to show the approximate positions of the central stars. The jet axes of the two components from \protect{\citet{Rodriguez2003}} are shown by the dotted white lines. The contour levels in the 93 GHz image are -3, 50, 100, 150, 200, 250, and 300 $\times\ \sigma_{rms}$, where $\sigma_{rms}$ is the root-mean-square noise of the image. For the 153 GHz and 225 GHz images, the contour levels are -3, 25, 50, 75, 100, 125, and 150 $\times\ \sigma_{rms}$. For the 336 GHz image, the contour levels are -3, 25, 50, 75, 100, and 125 $\times\ \sigma_{rms}$. The synthesised beam for each image is shown in the bottom left corner. The values for the synthesised beam and $\sigma_{rms}$ of each image are given in \protect{\autoref{tab:L1551_IRS_5_fluxes}}.}
    \label{fig:L1551_IRS_5_ALMA}
\end{figure*}

For all of the VLA and VLA+e-MERLIN bands, a Briggs weighting of $+0.5$ was used for imaging as this seemed to give the best compromise between angular resolution and sensitivity. Also, when imaging the combined e-MERLIN and VLA image, an outer uv-cutoff of $1.375 \times 10^6\ \lambda$ was used to show the extended emission more clearly. Note that the Ku Band and K Band data were each split into two sub-bands given their large fractional bandwidths in order to increase spectral sampling in the SED.

The total flux densities of the sources at each frequency were measured in CASA by drawing a region around each source containing all of the emission above a certain threshold  and integrating the flux within that region. The thresholds used were $3\sigma_{\mathrm{rms}}$ for the VLA and e-MERLIN images, $10\sigma_{\mathrm{rms}}$ for the ALMA Band 3 and Band 4 images, and $25\sigma_{\mathrm{rms}}$ for the ALMA Band 6 and Band 7 images. Higher thresholds were used for the ALMA images to avoid including dust emission from the circumbinary rings around the sources. For some of the images, particularly in the C Band image of L1551 IRS 5, there is overlap between the emission of the two sources. In these cases, the polygon was drawn so that it passes through the saddle-point of the brightness distribution between the two sources. Although this mean that there is a significant degree of uncertainty in the flux density of the N component of L1551 IRS 5 at 5 GHz, due to the low level of its flux density and the high degree of overlap with the emission of the S component.

\begin{figure*}[h!]
\centering
    \includegraphics{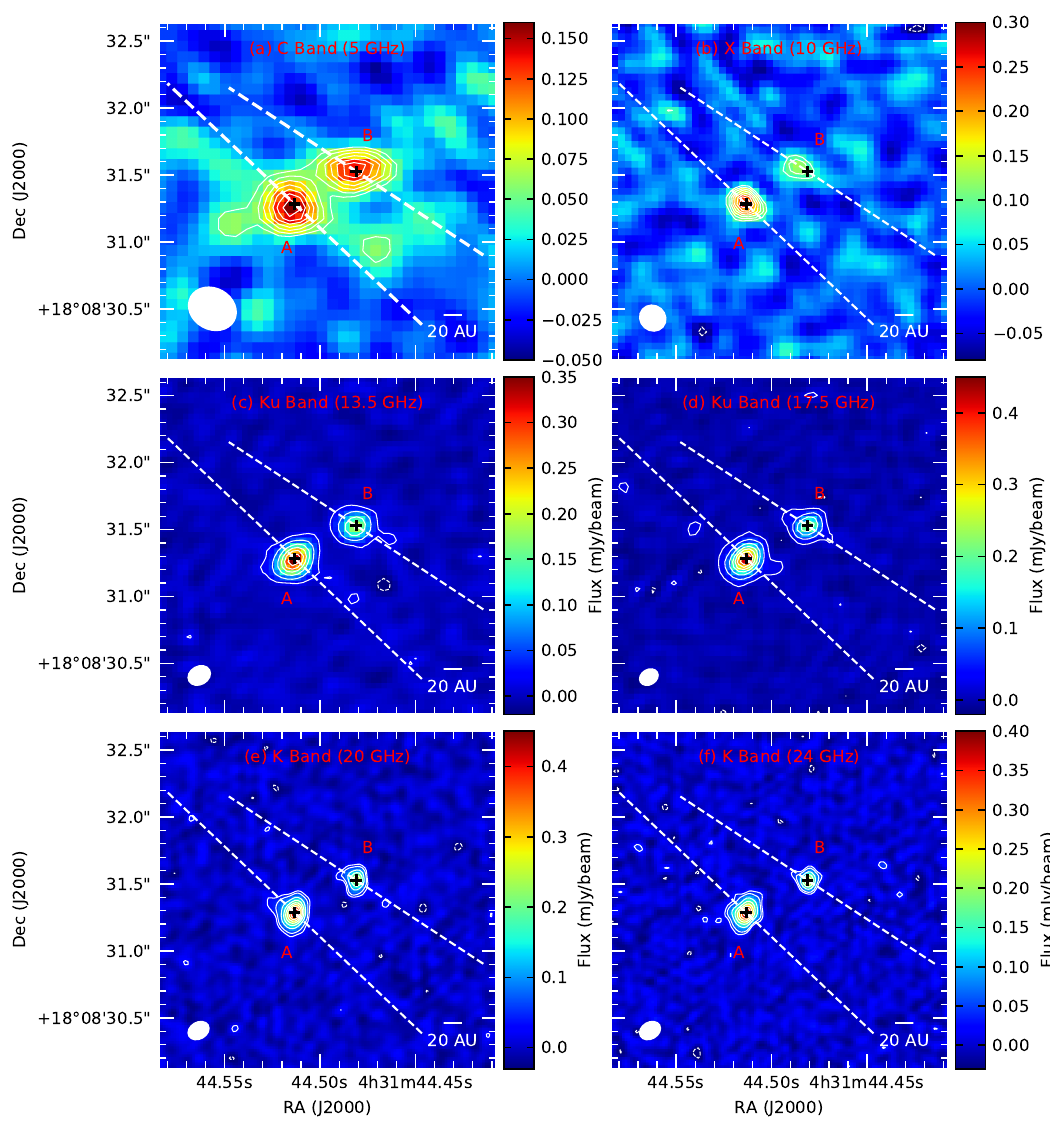}
    \caption{Images of L1551 NE from the VLA observations (contours + colour scale) as follows: (a) the VLA 6 GHz image. (b) the VLA 10 GHz image (c) the VLA 13.5 GHz image (d) the VLA 17.5 GHz image (e) the VLA 20 GHz image (f) the VLA 24 GHz image. The black crosses in each image indicate the peaks of the disk emission of the two sources in the 24 GHz image to show the approximate positions of the central stars. The jet axes of the two components from \protect{\citet{Reipurth2002}} are shown by the dotted white lines. The contour levels in the 6 GHz image are -3, 3, 4, 5, 6, 7, 8, and 9 $\times\ \sigma_{rms}$, where $\sigma_{rms}$ is the root-mean-square noise of the image. For the 10 GHz image, the contour levels are -3, 3, 4, 5, 6, 7, 8, 9, and 10 $\times\ \sigma_{rms}$. For the 13.5 GHz and 17.5 GHz images, the contour levels are -3, 3, 10, 20, 30, 40, 50, and 60 $\times\ \sigma_{rms}$. For the 20 GHz image, the contour levels are -3, 3, 5, 10, 15, 20, 25, 30, 35, and 40$\times\ \sigma_{rms}$. For the 24 GHz image, the contour levels are -3, 3, 5, 10, 15, 20, 25, 30, and 35 $\times\ \sigma_{rms}$. The synthesised beam for each image is shown in the bottom left corner. The values for the synthesised beam and $\sigma_{rms}$ of each image are given in \protect{\autoref{tab:L1551_NE_fluxes}}.}
    \label{fig:L1551_NE_C_to_K_Band}
\end{figure*}

\begin{figure*}
\centering
    \includegraphics{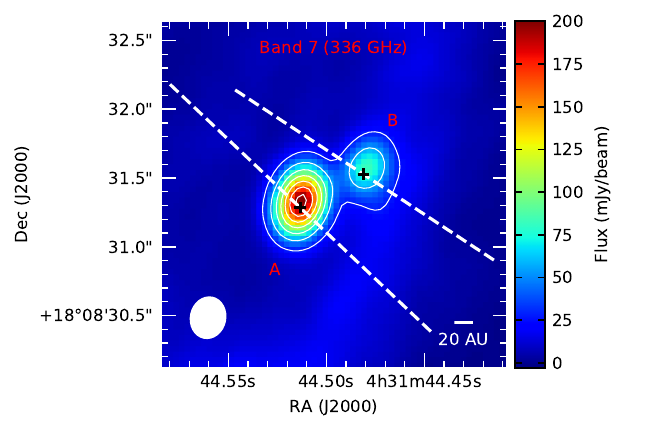}
    \caption{Image of L1551 NE from the 336 GHz ALMA observation (contours + colour scale). The black crosses in the image indicate the peaks of the disk emission of the two sources in the 24 GHz image to show the approximate positions of the central stars. The jet axes of the two components from \protect{\citet{Reipurth2002}} are shown by the dotted white lines. Note that there is a displacement between the peaks of the ALMA 336 GHz image and the 24 GHz image due to proper motion of the source between the epochs (Oct 2019 and Jan 2021) which was not corrected for in the image. The contour levels are -3, 25, 50, 75, 100, 125, 150, and 175 $\times\ \sigma_{rms}$, where $\sigma_{rms}$ is the root-mean-square noise of the image. The synthesised beam for the image is shown in the bottom left corner. The values for the synthesised beam and $\sigma_{rms}$ are given in \protect{\autoref{tab:L1551_NE_fluxes}}.}
    \label{fig:L1551_NE_ALMA_Band_7}
\end{figure*}

When measuring the total flux densities of the sources in the e-MERLIN and VLA observations, the data at each frequency was re-imaged, using the same outer uv-cutoff and synthesised beam for each band. This was in order to measure and compare the total flux densities at different frequencies more accurately and avoid any issues with differences in uv-coverage and synthesised beam size between the images. The uv-cutoff and synthesised beam used was the maximum baseline and the synthesised beam of the lowest resolution observation. For L1551 IRS 5, this was the X Band VLA observation, which gives a uv-cutoff of $1.375 \times 10^6\ \lambda$ and a beam of $0\farcs20 \times 0\farcs18$ with a PA of $49\degr$. For L1551 NE, this was the C Band VLA observation, which gives a uv cut-off of $1\times 10^6\ \lambda$ and a beam of $0\farcs37 \times 0\farcs30$ with a PA of $61\degr$. Note however, that the images presented in \autoref{fig:L1551_IRS_5_C_to_K_Band} - \autoref{fig:L1551_NE_ALMA_Band_7}, are the original images before all of the frequency bands were re-imaged with the same synthesised beam and uv-cutoff. Also note that the synthesised beam values given in \autoref{tab:L1551_IRS_5_fluxes} and \autoref{tab:L1551_NE_fluxes}, are the synthesised beams in the original images.

\section{Results}
\label{sec:results}
\subsection{L1551 IRS 5}

L1551 IRS 5 was imaged and successfully detected in all bands. The flux densities measured for the N and S components at each frequency are listed in \autoref{tab:L1551_IRS_5_fluxes}. The images from the VLA observations are shown in \autoref{fig:L1551_IRS_5_C_to_K_Band} in colour scale, with contours overlaid. The black crosses indicate the peaks of the disk emission from the ALMA Band 4 image (153 GHz/2.0 mm) to show the approximate positions of the central stars. The peaks in the ALMA Band 4 image were used as this was the highest resolution image obtained. These positions have been shifted, relative to the ALMA image, to account for the proper motion between the ALMA observation and the VLA observations based on the proper motion measurements from \citet{Villa2017}. The dashed white lines indicate the position angles of the jets from \citet{Rodriguez2003}.

The combined e-MERLIN and VLA C Band (5 GHz/6 cm) image is shown in \autoref{fig:L1551_IRS_5_C_to_K_Band} (a). At this frequency, it is expected that there should be very little dust emission from the disk and thermal free-free emission from the jet should be dominant. This is consistent with what is seen in the S component, where the emission is clearly elongated along the jet axis. Two main emission lobes are seen either side of the central star, with the NE lobe corresponding to the red-shifted jet and the SW lobe corresponding to the blue-shifted jet. In the N component on the other hand, the emission detected appears to be very weak and close to the position of the central star with no extended emission seen along the jet axis.

At the higher frequencies observed with the VLA (10 - 24 GHz), the emission of both sources peaks at the positions of the central stars, with extended emission seen along the direction of the jet axes of both sources. The double lobe structure seen at 5 GHz in the S component is not seen at higher frequencies. This is likely due to lower angular resolution along the jet axis as well as decreasing optical depth in the jet at higher frequencies causing the peaks of the free-free emission to be closer to the origin of the jet.

At ALMA frequencies (93 - 336 GHz), the emission is expected to be completely dominated by dust emission, with little or no free-free emission present. The ALMA images shown in \autoref{fig:L1551_IRS_5_ALMA} agree with this, as the emission is centred at the positions of the stars at all frequencies, with no emission seen along the jet axes.

\subsection{L1551 NE}
L1551 NE was successfully imaged and detected in all of the bands in which it was observed. The flux densities were measured for component A and component B at each frequency using the CASA task \emph{imfit} and are shown in \autoref{tab:L1551_NE_fluxes}.

The images from the VLA and ALMA observations are shown in \autoref{fig:L1551_NE_C_to_K_Band} and \autoref{fig:L1551_NE_ALMA_Band_7} in colour scale, with contours overlaid. To show the approximate positions of the central stars, the peaks of the emission from the VLA 24 GHz image are indicated by the black crosses. The peaks of the 24 GHz image were used as this was the highest resolution image obtained. Unlike in the case of L1551 IRS 5, the peak in the dust emission seen in the ALMA 336 GHz image was not used as there was proper motion between the epochs of the ALMA observation (Oct 2019) and the VLA observations (Dec 2019 - Jan 2020), which was not corrected for, as no proper motion measurements were available for L1551 NE. This can be seen in the displacement between the peaks of the ALMA Band 7 image and the 24 GHz image in \autoref{fig:L1551_NE_ALMA_Band_7}.

The dashed white lines indicate the position angles of the jets from \citet{Reipurth2002}. Similar to L1551 IRS 5, we should expect to see mainly free-free emission at lower frequencies, while at higher frequencies dust emission from the disk should become dominant.

\section{Discussion}
\label{sec:discussion}

\subsection{Spectrum}
\begin{figure*}
\centering
    \includegraphics{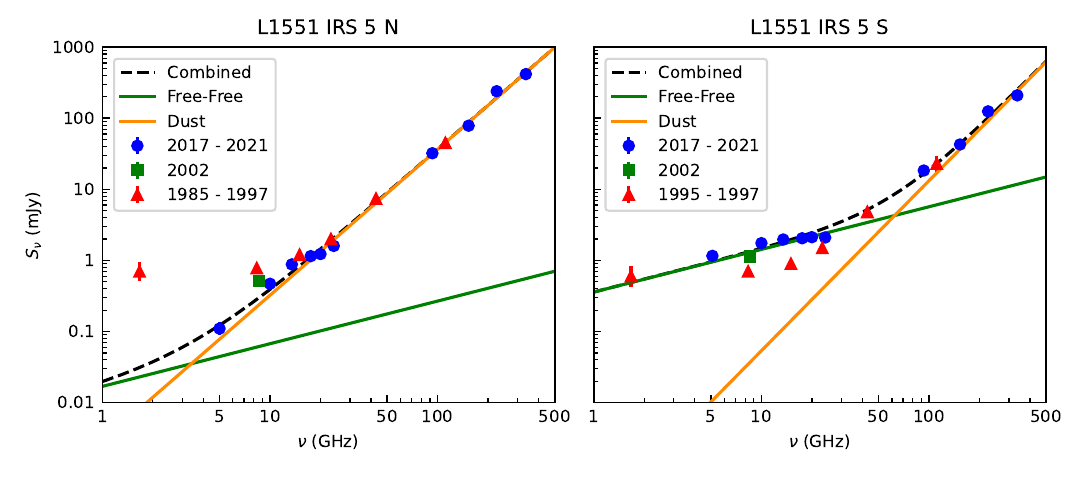}
    \caption[Spectra of the N and S components of L1551 IRS 5]{The spectra of the N component (left) and the S component (right) of L1551 IRS 5 including the total flux density values from this work obtained in 2017 - 2021 (blue circles), from \protect{\citet{Rodriguez2003}} obtained in 2002 (green square), and from \protect{\citet{Rodriguez1998}} obtained in 1985 - 1997 (red triangles). The spectra of the 2017 - 2021 data in both panels were fitted with a combination of free-free emission from the jet (green line) and dust emission from the disk (orange line). The combined fits are shown by the dashed black lines.}
    \label{fig:L1551_IRS_5_N+S_spectrum}
\end{figure*}

To learn more about the emission observed from these sources, the total flux densities of the components of L1551 IRS 5 and L1551 NE were plotted as a function of frequency to give a spectrum for each source. The resulting spectra for the N and S components of L1551 IRS 5 are shown in \autoref{fig:L1551_IRS_5_N+S_spectrum}. For comparison, the flux density values given by \citet{Rodriguez1998} and \citep{Rodriguez2003} are shown in the figures as well. Both components were fitted with a model spectrum consisting of a combination of a thermal free-free emission component and a dust emission component. When modelling the free-free emission, the model from \citet{Reynolds1986} was used. This assumes that the ionised flow in the jet begins at an injection radius above the disk along the jet axis of $r_0$, with an initial velocity $v_0$, electron temperature $T_0$, half-width $w_0$, and electron density $n_0$. The half-width and electron density of the jet vary with distance along the jet as $w(r) = w_0 (r/r_0)^{\epsilon}$ and $n_e(r) =n_0(r/r_0)^{q_n}$ respectively. A conical jet model was used for which $\epsilon=1$, $q_n = -2 \epsilon = -2$ and the electron temperature and velocity of the jet is constant ($T_e(r) = T_0$, $v(r) = v_0$). The spectrum model for the thermal free-free emission is then:
\begin{align}
\begin{split}
    \left( \frac{S_{\nu}}{\mathrm{mJy}} \right) = & 1.9 \times 10^{-7} \left( \frac{D}{\mathrm{pc}} \right)^{-2} \left(\frac{T_{\mathrm{e}}}{\mathrm{K}} \right)^{\mathbf{1/10}} \left( \frac{r_0}{\mathrm{au}} \right) \left( \frac{w_0}{\mathrm{au}} \right)^{5/3} \left( \frac{n_0}{\mathrm{cm^{-3}}} \right)^{4/3} \\ 
    & \times(\sin i)^{1/3} \left( \frac{\nu}{\mathrm{GHz}} \right)^{0.6}\\
    = & K_{\mathrm{ff}} \left( \frac{\nu}{\mathrm{GHz}} \right)^{0.6}
\end{split}
\end{align}
where $D$ is the distance to the object and $i$ is the inclination angle of the jet to the line of sight. When fitting the spectrum, all of the non-frequency parameters were combined into the parameter $K_{\mathrm{ff}}$ as a constant of proportionality for the free-free emission. Note that this equation is only accurate below the threshold frequency $\nu_\mathrm{m}$ above which the jet becomes entirely optically thin. However, for typical parameters of a jet this is usually estimated to be $\nu_{\mathrm{m}} > 40\ \mathrm{GHz}$ \citep{Anglada2018}, at which point the spectra of the sources are clearly dominated by dust emission.

The dust emission component is modelled using a simple power-law $S_{\nu} \propto \nu^{\alpha}$ giving a total model spectrum of:
\begin{equation}
\begin{split}
    \left( \frac{S_{\nu}}{\mathrm{mJy}} \right) = K_{\mathrm{ff}} \left( \frac{\nu}{\mathrm{GHz}} \right)^{0.6} + K_{\mathrm{dust}} \left( \frac{\nu}{\mathrm{GHz}} \right)^{\alpha_{\mathrm{dust}}}
\end{split}
\end{equation}
where $K_{\mathrm{dust}}$ is a constant of proportionality for the dust emission component and $\alpha_{\mathrm{dust}}$ is the spectral index of the dust emission.

The S component is fit very well by this, as seen in \autoref{fig:L1551_IRS_5_N+S_spectrum}, with the flux density at low frequencies being mainly due to free-free emission and at higher frequencies being mainly due to dust emission with a spectral index for the dust emission of $\alpha_{\mathrm{dust}} = 2.4\pm0.3$.

In the N component, the data is fit well by the model, as seen in \autoref{fig:L1551_IRS_5_N+S_spectrum}. However, the fit seems to suggest that there is almost no free-free emission present and the emission is almost entirely due to dust emission with a spectral index of $\alpha_{\mathrm{dust}} = 2.05\pm0.07$. Although it should be noted that in the Ku Band images of L1551 IRS 5 shown in \autoref{fig:L1551_IRS_5_C_to_K_Band}, there is clearly some emission seen along the jet axis of the N component, indicating that there is at least some level of free-free emission from the jet present.

\begin{figure*}
	\centering
	\includegraphics{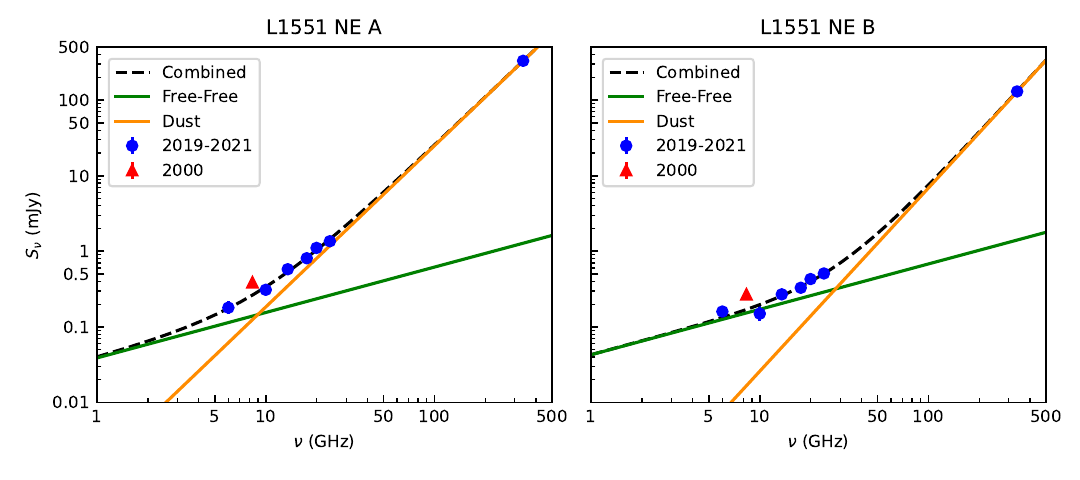}
	\caption[Spectrum of components A and B of L1551 NE]{The spectra of component A (left) and component B (right) of L1551 NE including the total flux density values from this work obtained in 2019 - 2021 (blue circles) and the values from \protect{\citet{Reipurth2002}} obtained in 2000 (red triangle). The spectra of the 2019 - 2021 data in both panels were fitted with a combination of free-free emission from the jet (green line) and dust emission from the disk (orange line). The combined fits are shown by the dotted black lines.}
	\label{fig:L1551_NE_A+B_spectrum}
\end{figure*}

For L1551 NE, we do not have as many data points as for L1551 IRS 5, particularly at higher frequencies where we only have ALMA observations in one band. However, it can still be seen in \autoref{fig:L1551_NE_A+B_spectrum} that the spectra for both components A and B can be fit well by a combination of thermal free-free emission and dust emission, with free-free emission dominant at lower frequencies and dust emission dominant at higher frequencies. The spectral indices for the dust component are $\alpha_{\mathrm{dust}} = 2.13 \pm 0.04$ and $\alpha_{\mathrm{dust}} = 2.4 \pm 0.2$ for components A and B respectively.

\subsection{Morphology}
When comparing the C Band image of L1551 IRS 5 shown in \ref{fig:L1551_IRS_5_C_to_K_Band} (a) with previous images at cm-wavelengths from \citet{Rodriguez1998} and \citet{Rodriguez2003}, it is clear that there are significant changes in the morphology of the jets of both sources. For example, there are two strong emission knots seen along the jet axis of the N component in the image from \citet{Rodriguez2003}, one to the NE and one to the SW. However, these seem to have disappeared in the images that we obtained. This is not that surprising when considering the proper motion of the material in the jet. \citet{Anglada2018} estimate a jet velocity of $v_{\mathrm{jet}} \sim 150\ \mathrm{km\ s^{-1}}$ for L1551 IRS 5. This would give a proper motion of $\sim 150\ \mathrm{mas\ yr^{-1}}$ at a distance of 147 pc with an inclination of $i = 45 \degr$. Since the \cite{Rodriguez2003} observations were taken in February 2002, this would give $\sim 2.6\arcsec$, or $\sim 400\ \mathrm{au}$, of proper motion between the two epochs. In addition, the emission from any ejected material is likely to decline with time as it moves outwards along the jet due to recombination of the ionised material and decreasing electron density. The recombination time for the ionised material in the jet is given by $t = 1/(n_e \alpha)$, where $\alpha$ is the recombination rate. For hydrogen, the recombination rate at $T=10^4\ \mathrm{K}$ is typically $\alpha = 4.16 \times 10^{-13} s^{-1} cm^3$ \citep{Pradhan2011}. Consider an emission knot in the jet of the N component at a distance of $\sim 100\ \mathrm{au}$ from the star. From the ionised mass-loss rate measured in \autoref{sec:ionised_mass_loss_rate} and $q_n =  - 2$ we can estimate an electron density of $n_e \sim 10^4\ \mathrm{cm^{-3}}$ at this distance in the jet. This would imply a recombination time of $\sim 7.6\ \mathrm{yrs}$. From this, it is clear that most of the ionised material in any emission knot can recombine between the two epochs and therefore the thermal emission would disappear.

One thing of interest to note is that in \citet{Rodriguez2003}, they detect a region of emission between the two jets $\sim 0.6\arcsec$ to the SW of the two sources. They speculate that this emission could be due to a region of shock interaction between the outer parts of the outflows. They state that to test this hypothesis, a lack of proper motion would have to be observed in this emission as an interaction zone between the two outflows should remain stationary. In \autoref{fig:L1551_IRS_5_C_to_K_Band} (a) and (b), a region of emission is also detected between the outflows to the SW of the two sources at a similar position to that of the emission detected by \citet{Rodriguez2003}. If this emission corresponds to the same emission region as the one they detected, this would indicate a lack of proper motion and back up their hypothesis of this being a region of shock interaction between the two outflows.

\begin{table*}
    \centering
    \caption{Ionised mass-loss rates of L1551 IRS 5 and L1551 NE radio jets}
    \begin{tabular}{c c c c c c c c c}
	\hline
	Source          & Epoch         & $K_{\mathrm{ff}}$ & $\theta_0$ &  $v$                    & $D$  & $T_{\mathrm{e}}$ & $i$ & $\dot{M}_{\mathrm{ion}}$ \\
                    &               & (mJy)             &            & ($\mathrm{km\ s^{-1}}$) & (pc) & (K)              &     & ($\mathrm{M_{\odot}\ yr^{-1}}$) \\
	\hline
	L1551 IRS 5 N   & 1985 - 1997   & $0.29\pm0.09$     & 0.75       & 150                     & 147  & $10^4$           & 45  & $(5.3 \pm 1.2) \times 10^{-9}$ \\
	                & 2017 - 2021   & $0.017\pm0.013$   & 0.75       & 150                     & 147  & $10^4$           & 45  & $(0.6 \pm 0.4) \times 10^{-9}$    \\
    \hline
    L1551 IRS 5 S   & 1985 - 1997   & $0.24\pm0.07$     & 0.49       & 150                     & 147  & $10^4$           & 45  & $(3.4 \pm 0.8) \times 10^{-9}$     \\
                    & 2017 - 2021   & $0.36\pm0.04$     & 0.49       & 150                     & 147  & $10^4$           & 45  & $(4.5 \pm 0.4) \times 10^{-9}$  \\
    \hline
    L1551 NE A      & 2019 - 2021   & $0.039\pm0.006$   & 0.52       & 160                     & 147  & $10^4$           & 58  & $(1.0 \pm 0.1) \times 10^{-9}$ \\
    \hline
	L1551 NE B      & 2019 - 2021   & $0.043\pm0.007$   & 0.52       & 70                     & 147  & $10^4$           & 58  & $(0.5 \pm 0.1) \times 10^{-9}$      \\
	\hline
    \end{tabular}
    \label{tab:ion_mass-loss_rates}
\end{table*}

\subsection{Flux Variability}
Variability in the flux density of L1551 IRS 5 can clearly be seen looking at the spectra in \autoref{fig:L1551_IRS_5_N+S_spectrum} and comparing the flux densities in this work with those from \citet{Rodriguez1998}, observed between 1985 - 1997, and from \citep{Rodriguez2003}, observed in 2002. It is clear that there is a high degree of variability in the jet emission of both sources, as the flux density at lower frequencies ($\nu \lesssim 20\ \mathrm{GHz}$), where the jet emission dominates, varies significantly between the three epochs. On the other hand, the flux density at higher frequencies ($\nu \gtrsim 20\ \mathrm{GHz}$), where the disk emission dominates, is similar at both epochs, indicating that the disk emission has not varied significantly. In the N component in particular, the flux density has decreased significantly by a factor of $\sim 5$ at 5 GHz compared with the data in \citet{Rodriguez1998}. This suggests that the free-free emission from the jet has drastically decreased between the two epochs. The S component on the other hand, shows more modest although still significant variability between the two epochs with an increase in flux density by a factor of $\sim 2$ at 5 GHz. The synthesised beam sizes are similar in all of  the observations, with beam sizes of $0\farcs25 \times 0\farcs24$ and $0\farcs18 \times 0\farcs12$ in the \citet{Rodriguez1998} and \citet{Rodriguez2003} 8.3 GHz observations respectively, compared with $0\farcs20 \times 0\farcs18$ in the 5 GHz observation in this work, and so the differences in flux density are unlikely to be due to differences in the scales of emission being recovered.

For L1551 NE, the flux densities obtained from this work can be compared with the data from \citet{Reipurth2002}, observed in 2000, as shown in \autoref{fig:L1551_NE_A+B_spectrum}. The synthesised beam size is similar for both epochs, $0\farcs37 \times 0\farcs30$ in our observations vs $0\farcs34 \times 0\farcs27$ in \citet{Reipurth2002}, and so they should recover similar scales of emission. While \citet{Reipurth2002} only observed L1551 NE at one frequency, it can still be seen that the flux density at low frequencies seems to have decreased in both components at 8 GHz between 2000 and 2021, with the flux density in component B decreasing by a factor of $\sim 2$. This would indicate that the free-free emission from the jet has decreased in both components.

This level of flux variability, particularly for the N component of L1551 IRS 5, where the free-free emission seems to have almost disappeared, is quite remarkable. For most YSOs observed, the free-free emission shows only modest variability on the order of $10 - 20 \%$ \citep[e.g.][]{Rodriguez2008,Rodriguez2014,Carrasco-Gonzalez2012}. The variability observed in L1551 IRS 5 and L1551 NE suggests that binary systems could show more temporal changes in their ionised mass loss rate than single sources. However, a few cases of extreme variability have been seen previously in presumably single stars. For example, the flux density of the radio jet in B335 was found to increase by a factor of $\gtrsim 5$ between 1994 and 2000 \citep{Avila2001,Reipurth2002} while DG Tau A showed an increase in flux density by a factor of $\sim 2$ between 1994 and 1996 \citep{Rodriguez2012}. Interestingly, \citet{Galfalk2007} propose that B335 could be a binary system. Additional studies of variability and multiplicity are needed to investigate this issue.

\subsection{Ionised Mass-loss Rate}
\label{sec:ionised_mass_loss_rate}
From the fit to the spectra of the sources, the ionised mass-loss rates of the jets can be estimated. From \citet{Reynolds1986}, the ionised mass-loss rate for a conical jet is given by:
\begin{equation}
\begin{split}
    \left( \frac{\dot{M}_{\mathrm{ion}}}{\mathrm{M_{\odot}\ yr^{-1}}} \right) = & 1.23 \times 10^{-16} K_{\mathrm{ff}}^{3/4} \theta_0^{3/4} \left( \frac{v}{\mathrm{km\ s^{-1}}} \right) \left( \frac{D}{\mathrm{pc}} \right)^{3/2}\\
    & \times \left( \frac{T_{\mathrm{e}}}{\mathrm{K}} \right)^{-0.075} (\sin (i))^{-1/4}
\end{split}
\label{eqn:ionised_mass_loss_rate}
\end{equation}
where $\theta_0$ is the opening angle of the jet. In addition to getting the ionised mass-loss rate based on our data, we can fit the data from \citet{Rodriguez1998} with the same model to get the values for a different epoch and compare it with the epoch of this work.

The values obtained for the ionised mass-loss rates in each of the jets in each epoch are listed in \autoref{tab:ion_mass-loss_rates} along with the values of the parameters of \autoref{eqn:ionised_mass_loss_rate} used when deriving the values. For this work, an electron temperature of $T_{\mathrm{e}} \sim 10^4\ \mathrm{K}$ was assumed for all of the sources. It should be noted that while the mass-loss rate derived depends on the value of $T_{\mathrm{e}}$ assumed, changing the value of $T_{\mathrm{e}}$ does not significantly affect the result, as $\dot{M}_{\mathrm{ion}}$ is only weakly dependent on $T_{\mathrm{e}}$ ($\dot{M}_{\mathrm{ion}} \propto T_{\mathrm{e}}^{0.075}$). A rough estimate for the opening angle can be found from the deconvolved major and minor axis of the jet emission: $\theta_0 \approx 2 \tan^{-1} (\theta_{\mathrm{min}} / \theta_{\mathrm{maj}})$. For L1551 IRS 5, using the values of $\theta_{\mathrm{min}}$ and $\theta_{\mathrm{maj}}$ obtained by \citet{Rodriguez2003}, this gives estimates of $\theta_0 \approx 43\degr$ and $\theta_0 \approx 28\degr$ for the N and S component respectively. For L1551 NE, neither our observations nor the observations from \citet{Reipurth2002} have sufficient resolution to estimate the major and minor axes of the jet. Therefore, we use a reasonable estimate of $\theta_0 \approx 30 \degr$ for these jets. For the jet velocity, the value for L1551 IRS 5 from  \citet{Anglada2018} was used, while for L1551 NE, Eqn. 12 from \citet{Anglada2018} was used to obtain rough estimates of $v$ for the two components based on the stellar masses of the sources, which are $M_{\mathrm{A}} \sim 0.67 \mathrm{M_{\odot}}$ and $M_{\mathrm{B}} \sim 0.13 \mathrm{M_{\odot}}$ respectively \citep{Takakuwa2012,Takakuwa2014}.

For the N component of L1551 IRS 5, it can be seen that there has been a factor of $\sim 10$ decrease in the ionised mass-loss rate of the jet between 1998 and 2020. While in the S component, there has been a small increase in the ionised mass-loss rate of the jet between the two epochs. Such changes are of course to be expected based on the change in flux density. Note that while these changes in the ionised mass-loss rates are likely due to changes in the total mass-loss rate of the jets, they could also be due to changes in the ionisation fractions of the jets.

\begin{figure*}
	\centering
	\includegraphics{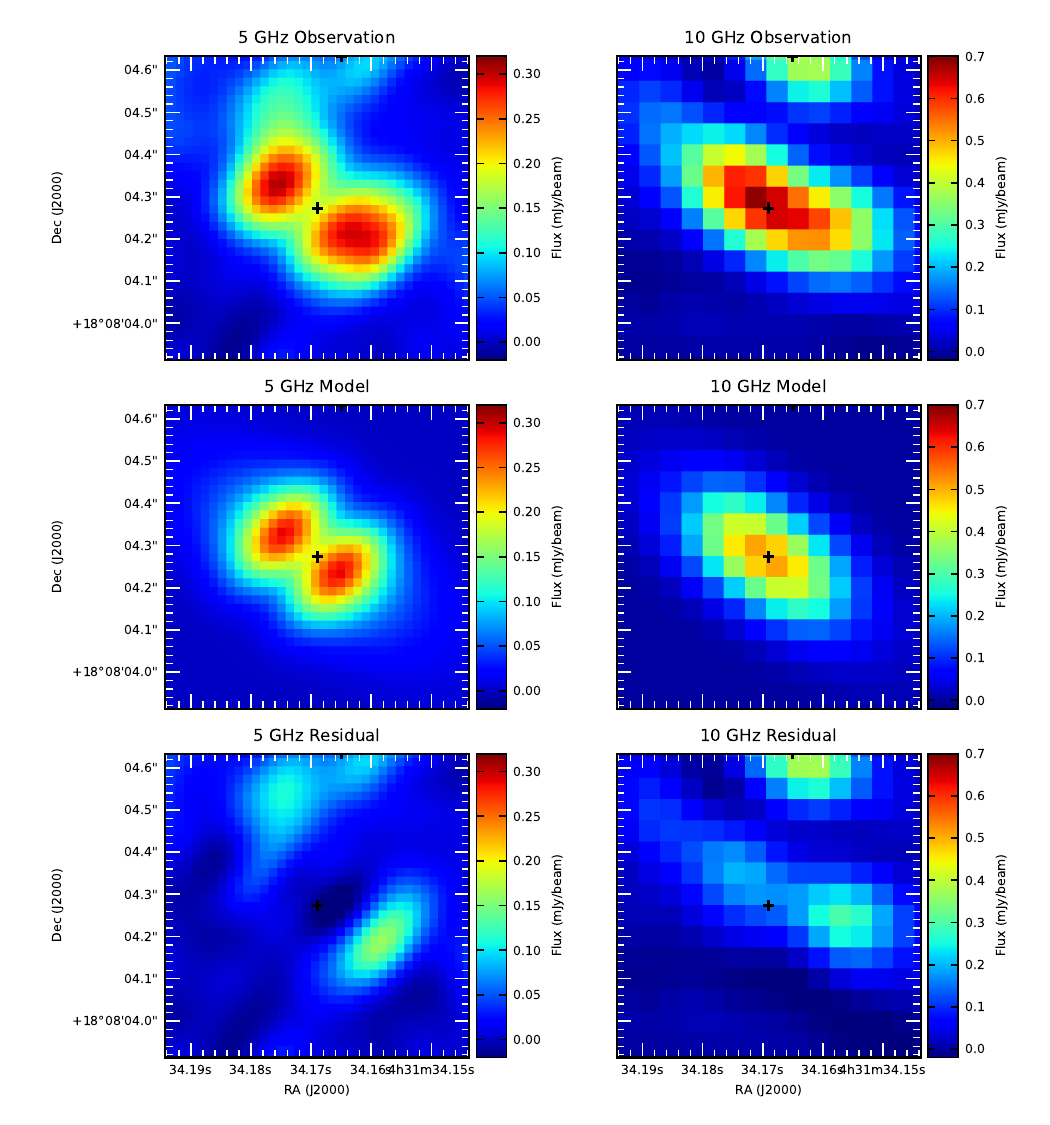}
	\caption{The VLA+e-MERLIN 5 GHz image (left) and the VLA 10 GHz image (right) of the S component of L1551 IRS 5 compared with the images obtained from radiative transfer modelling (middle) and with the residual images obtained by subtracting the model images from the observed images (bottom). The peak of the emission in the ALMA 153 GHz image is indicated by the black cross to show the approximate position of the central star.}
	\label{fig:simulated_image_comparison}
\end{figure*}

We can compare the values of $\dot{M}_{\mathrm{ion}}$ estimated with the total mass-loss rate of L1551 IRS 5 to obtain values for the ionisation fraction of the jets. Based on SOFIA observations of [\ion{O}{I}]$_{63}$ emission from 2019, \citet{Sperling2021} calculated the mass-loss rate of the atomic jet of L1551 IRS 5. They determine that the [\ion{O}{I}]$_{63}$ emission in L1551 IRS 5 can be described by the \citet{Hollenbach1989} shock model, which predicts that the [\ion{O}{I}]$_{63}$ emission is produced in dissociative J-shocks in the jet. Applying this model to L1551 IRS 5, they estimate, based on the luminosity of the [\ion{O}{I}]$_{63}$ emission, that the mass-loss rate is $4.9 -- 5.4 \times 10^{-7}\ \mathrm{M_{\sun}\ yr^{-1}}$. By comparing this with mass-loss rate estimates based on other atomic and molecular emission lines, they determine that the atomic component is the dominant component of the jet. Therefore, this estimate for the mass-loss rate of the atomic component should provide a good estimate for the total mass-loss rate of the jet of L1551 IRS 5. Unfortunately, the two individual jets are unresolved and so we only have the total mass-loss rate for the whole system. But using this value as an upper limit for the total mass-loss rate in each jet gives an estimate for the ionisation fraction in each jet of $\gtrsim 0.1\%$ and $\gtrsim 1\%$ for the N and S components respectively. This ionisation level can be readily produced by an X-wind model \citep[e.g.][]{Shang2004} with the mass loss rates inferred. However, these are only rough estimates for the ionisation fractions given the large errors involved.

For L1551 NE, we cannot calculate the previous ionised mass-loss rate from a fit to the spectrum as \citet{Reipurth2002} only observed the source at one frequency. However, from this measurement there appears to have been a decrease of a factor of $\sim 2$ in flux between 2000 and 2021 in the free-free portion of the spectra of the components. Based on this, we can infer that there has likely been a decrease in the free-free emission of the jets and therefore a decrease in the ionised mass-loss rates of both of the jets.

\subsection{Radiative Transfer Modelling}
In order to better understand the properties of the jet, radiative transfer modelling was performed for the jet in the S component of L1551 IRS 5 using the package \emph{sf3dmodels} \citep{Izquierdo2018} to model the jet and RADMC-3D \citep{Dullemond2012} to then perform the radiative transfer calculations. \emph{sf3dmodels} creates a three-dimensional grid of the temperature, ionisation fraction, and density of the jet based on the model of \citet{Reynolds1986}. This allows us to obtain estimates for the injection radius $r_0$, i.e. at what distance above the disk the ionised flow in the jet begins, as well as the ionised mass-loss rate $\dot{M}_{\mathrm{ion}}$. To create this model, values for $\dot{M}_{\mathrm{ion}}$, $r_0$, the velocity of the jet $v_0$, electron temperature $T_e$, half-width at the base of the jet $w_0$, and inclination angle $i$ of the jet are required. The model of an ionised, conical jet from \citet{Reynolds1986} was used again, with $\epsilon=1$, $q_n = -2 \epsilon = -2$, and a constant temperature and velocity in the jet. The jet was modelled out to a distance of 120 au from the star and the resolution of the 3-D grid was 0.25 au.

Once the physical models have been generated, RADMC-3D is used to solve the equation of radiative transfer for free-free emission to generate a synthetic image of the region at a given wavelength. For L1551 IRS 5, we simulated images at 5 GHz and 10 GHz, since at these frequencies almost all of the emission should be free-free emission as opposed to dust emission and so it should be possible to replicate the observed image using only free-free emission. The model images created were convolved with the same synthesised beam as the observed image of the same frequency in order to compare the two. We aimed to recreate the jet morphology, i. e. the double lobe structure in the 5 GHz image, and the flux density seen in the observed images as best we can. To assess the accuracy of the model images, residual images were also generated by subtracting the model images from the observed images.

When creating the model, an estimate for the jet velocity of $v_0 = 150\ \mathrm{km\ s^{-1}}$ and a temperature of $T_{\mathrm{e}} = 10^4\ \mathrm{K}$ was used. The half-width $w_0$ was taken to be given by $r_0$ and the opening angle of the jet $\theta_0 = 2 (w_0/r_0)$. The morphology of the jet in the image was then mainly adjusted using the input value for $r_0$, while the flux density was adjusted using the value for $\dot{M}_{\mathrm{ion}}$.

It was found that the observed jet was best reproduced using values of $r_0 \approx 10\ \mathrm{au}$ and $\dot{M}_{\mathrm{ion}} \approx 4.3 \times 10^{-9}\ \mathrm{M_{\odot}\ yr^{-1}}$. This can be seen and compared with the observed image in \autoref{fig:simulated_image_comparison}. Note that the ionised mass-loss rate obtained is nearly the same as that predicted by \autoref{eqn:ionised_mass_loss_rate}. It can be seen the model images reproduce the observed images quite well, as the double lobe structure in the 5 GHz image is observed and the flux densities are similar. This can also be seen in the residual images, as most of the observed emission has been subtracted. Although in both the 5 GHz and 10 GHz images, there is some emission to the SW of the source in the blue-shifted jet which is still present in the residuals. This is likely as the jet is asymmetric, which was not accounted for in the model.

The models obtained for the 5 GHz and 10 GHz emission both predict that the free-free emission flux density should consist of two peaks on either side of the central star, with the emission then declining along the jet axis. However, when the images are convolved with the synthesised beams of the observed images, it can be seen that while in the 5 GHz image two peaks are seen, in the 10 GHz image the two peaks are not resolved. Instead, the emission is elongated with a peak at the position of the central star. This is due mainly to the lower angular resolution along the jet axis in the 10 GHz image. In addition, the optical depth at higher frequencies is lower and so the peaks of the emission are closer together.

The value obtained for $r_0$ could be interpreted as being due to variability in the mass loss rate of the jet. If the two lobes represent a recent increase in the ionised mass-loss rate followed by a decrease, this would explain the lack of emission closer to the origin of the jet. This would also be consistent with knots along the jet being due to variations in mass-loss-rate. For $r_0 \approx 10\ \mathrm{au}$ and a velocity of $v_0 = 150\ \mathrm{km\ s^{-1}}$, this would indicate variations in the mass-loss rate of the jet on timescales of $\sim$  4 months. Alternatively, it could be the case that the material in the jet is only ionised at a distance of 10 au, possibly due to a shock in the jet, or a new ionising event due to variability. As a result, there would be no ionised material closer to the origin of the jet.

These results have implications for the study of the jet launching mechanism, as any model of jet generation such as the X-wind model or the disk wind model, would need to explain the double lobe structure observed and the value of $r_0$ obtained for L1551 IRS 5 S. For example, through the jet being ionised at a shock front at a distance of $r_0$ or through mass-loss rate variability on timescales of a few months. In magnetocentrifugal wind models, the Alfv\'en radii are much smaller than where the jet self-collimates. The strong ionisation and high density from the jet axis out to the jet half-width $w_0$ provide constraints on the wind launching mechanism. In particular, the observed jet velocity, high ionisation, and the inferred high ionised jet density from the axis to $w_0$ are normal characteristics of the X-wind model, where the wind is launched from a radius much smaller than $w_0$ \citep[e.g.][]{Shang2002,Shang2004}. For L1551 IRS 5 S, the jet width $w_0$ is found to be $w_0=0.5 \theta_0 r_0$. Given $\theta_0=0.49$, and $r_0=10$ au, $w_0$ is shown to be $2.5$ au. Furthermore, the high jet velocity of $v_0 \sim 150$ km/s implies a launch radius $\lesssim 0.2$ au, which is the innermost region of the disk.

Finally, our results also show the importance of high-resolution observations of free-free emission from radio jets, particularly at lower frequencies ($\sim5\ \mathrm{GHz}$) where dust emission is negligible. By comparing the observations with simulations, we were able to probe the jet launching and collimation regions.

\section{Conclusions}
\label{sec:conclusions}
L1551 IRS 5 and L1551 NE were successfully imaged at a wide range of frequencies using high spatial resolution observations with e-MERLIN, the VLA, and ALMA. In particular, the 5 GHz image of L1551 IRS 5 that was obtained is one of the highest resolution images of a radio jet from a low-mass YSO obtained to date, with emission resolved on a scale of only $\sim 15\ \mathrm{au}$. Emission was detected from all of the components of the two binary systems at all frequencies. By plotting and fitting the spectra of the sources, it is apparent that free-free emission dominates at lower frequencies while dust emission dominates at higher frequencies, as expected. Although for the N component of L1551 IRS 5, there appears to be very little free-free emission present.

Significant changes in morphology and high levels of variability in flux density were seen in both components of L1551 IRS 5. In particular, the N component of L1551 IRS 5 appears to decline in flux density by a factor of $\sim 5$ at lower frequencies, suggesting that the free-free emission of this source has almost disappeared.

From fitting the spectra, estimates for the ionised mass-loss rates of each source were derived. By comparing the values obtained to an estimate for the total mass-loss rate based on atomic line observations, it was estimated that the ionisation fractions of the jets in L1551 IRS 5 are on the order of $\gtrsim 0.1\%$ and $\gtrsim 1\%$ for the N and S components respectively.

Given the variability in flux density of the free-free emission of the jets, this implied there is variability in their ionised mass-loss rates. By comparing the mass-loss rates derived based on the data in this work and from a previous epoch, it was shown that there is significant variation in the ionised mass-loss rate of the jets, particularly the N component of L1551 IRS 5, on timescales of several years. This indicates that there has been either significant changes in the total mass-loss rates of the jets or alternatively,  changes in the ionisation fraction of the jets.

The high resolution and sensitivity achieved in the 5 GHz image of L1551 IRS 5 made it possible to study the inner region of the radio jet of the S component in detail. By creating a physical model and then using radiative transfer modelling to create a simulated image which could be compared with the observed image, it was possible to obtain an estimate for the injection radius and ionised mass-loss rate of the jet of $r_0 \approx 10\ \mathrm{au}$ and $\dot{M}_{\mathrm{ion}} \approx 5.7 \times 10^{-9}\ \mathrm{M_{\odot}\ yr^{-1}}$ respectively. The strong ionisation and high density from the jet axis to its width $w_0\approx 2.5$ au, support the normal characteristics of an X-wind model. The high jet velocity of $v_0 \sim 150$ km/s suggests that the underlying magnetocentrifugal wind comes from the innermost $0.2$ au or smaller. It is very difficult to produce these findings by a disk wind. Our data and the modelling analysis thus favour the X-wind model.

These results show the use of high-resolution, high-sensitivity radio/mm observations for the study of radio jets from YSOs. Future ultra-sensitive interferometers such as the Square Kilometer Array (SKA) or the Next-Generation Very Large Array (ngVLA) will allow for many more YSOs to be studied in this way and potentially help us learn more about the innermost regions of protostellar jets.

\begin{acknowledgements}
AFJ, SJDP and TPR acknowledge funding from the European Research Council (ERC) under Advanced Grant No.\ 743029. AFJ also acknowledges support from NAOJ ALMA Scientific Research Grant code 2019-13B. C.C.-G. and R.G.M. acknowledge support from UNAM DGAPA-PAPIIT grants IN108822 and IG101321 as well as from CONACyT Ciencia de Frontera project ID 86372. A.R.-K. thanks the UNAM DGAPA Postdoctoral Fellowship Program. S.L. acknowledges support from UNAM-PAPIIT grant IN103921.
\end{acknowledgements}

%
%

\bibliographystyle{aa}
\bibliography{L1551}

\end{document}